\documentclass[a4paper,fleqn]{cas-sc}

\usepackage[numbers,sort&compress]{natbib}
\usepackage[utf8]{inputenc}
\DeclareUnicodeCharacter{2212}{\textminus}
\usepackage{textgreek}
\DeclareUnicodeCharacter{03B4}{\textdelta}
\usepackage{bbding}
\usepackage{subcaption}
\usepackage{cite}
\usepackage{amsmath,amssymb,amsfonts}
\usepackage{algorithmic}
\usepackage{graphicx}
\usepackage{textcomp}
\usepackage{xcolor}
\usepackage{cuted}
\usepackage{float}
\usepackage{nomencl}
\usepackage{framed}
\usepackage{multicol}
\usepackage{threeparttable}

\def\tsc#1{\csdef{#1}{\textsc{\lowercase{#1}}\xspace}}
\tsc{WGM}
\tsc{QE}

\begin{document}
\let\WriteBookmarks\relax
\def\floatpagepagefraction{1}
\def\textpagefraction{.001}
\ExplSyntaxOn
\cs_gset:Npn \__first_footerline: {}
\ExplSyntaxOff

\shorttitle{Inverse thermal modeling}    

\shortauthors{G. Fu et~al.}  

\title [mode = title]{Fast Forward and Inverse Thermal Modeling for Parameter Estimation of Multi-Layer Composites - Part II: Inverse Modeling and Applications}  



%

\author{Gan Fu}[orcid=0009-0007-6899-1209]
\author{Mitrofan Curti}
\author{Calina Ciuhu}
\author{Elena A. Lomonova}


\ead{g.fu@tue.nl}



\affiliation{organization={Department of Electrical Engineering, Eindhoven University of Technology, 5612 AZ Eindhoven, The Netherlands}}






\cortext[1]{Corresponding author}



\begin{abstract}
A fast inverse heat conduction model (IHCM) is developed for estimating unknown properties of multi-layer composites considering internal heat generation. This work builds on the validated analytical forward models presented in Part I. Transient temperature at a single point is used as input, with the objective function minimized through an interior-point optimization algorithm. The IHCM accurately estimates thermal properties such as thermal conductivity, specific heat capacity, density, and heat transfer coefficient. It also identifies internal geometric variations and their locations, such as delamination caused by thermal expansion or mechanical motion. These predictions are validated through finite element (FE) simulations. Additionally, a sensorless strategy is introduced, providing a non-invasive inverse modeling approach. The feasibility, sensitivity and limitations of the proposed IHCM are evaluated across various scenarios. The results demonstrate strong potential for applications such as thermal performance monitoring, online defect detection, and real-time diagnostics in multi-layer composite systems.

\end{abstract}





\maketitle










\section{Introduction}

Understanding the thermal behavior of multi-layer composite structures is essential in many engineering applications~\citep{demonteTransientHeatConduction2000}. However, factors such as mechanical stress, prolonged operation, and manufacturing tolerances introduce deviations in material properties and internal geometry. These deviations may reflect underlying defects such as cracks, delamination, and thermal aging.
As a result, such variations lead to uncertainties that can impact the accuracy of thermal performance evaluations using either analytical or numerical methods.
To account for these uncertainties, the inverse heat conduction problem (IHCP) is widely used. By incorporating appropriate temperature or heat flux measurements, IHCPs enable the estimation of unknown quantities within the system~\citep{alifanovInverseHeatTransfer1994,colacoInverseOptimizationProblems2006,woodburyInverseHeatConduction2023a}.

An inverse problem is using the measurements of the system or process state, to specify unknown characteristics causing this state~\citep{alifanovInverseHeatTransfer1994}. Unlike the well-posed forward problems, IHCPs are typically ill-posed, as they may violate one or more of the conditions for the existence, uniqueness, or stability of the solution~\citep{hadamard1923lectures}. IHCPs may be formulated either as function estimation~\citep{woodburyInverseHeatConduction2023a} or parameter estimation problems~\citep{beck1977parameter,beck1998inverseproblems}, with the latter being the main subject of this paper.

IHCPs for multi-layer composites have been applied in various domains. For instance, emissivity and thermal conductance of multi-layer thermal insulation blankets in spacecraft are estimated using heat flux and temperature measurements~\citep{alifanovStudyMultilayerThermal2009}. Position-dependent thermal conductivity and heat capacity have been identified in layered systems~\citep{chengSurfaceHeatFlux2025}. Surface temperature and flux are predicted in a two-layer inverse model for thermal analysis of aircraft skins, with the temperature solution derived by Green's function method~\citep{avilesExactSolutionComposite1998}. A broad comparison of IHCP methods of finding the surface heat flux using transient temperature measurements inside a heat-conducting body is available in~\citep{beckComparisonInverseHeat1996}. Other studies estimate unknown heat flux in cylindrical coordinates~\citep{chuInverseProblemsAxisymmetric2003} and Cartesian coordinates~~\citep{guoReverseIdentificationMethod2021}, while accounting for interfacial thermal resistance. Thermal conductivity has been estimated in a multi-layer thermal protection system for high-temperature fuel cells, using numerical forward model that combines radiation and conduction for porous materials based on finite volume method~\citep{huangCalculationHightemperatureInsulation2014}. Inverse problem has been also used to estimate thermal conductivity in a three-layer wall building system~\citep{jumabekovaOptimalObservationSequence2020}. 
Multi-layer composite IHCPs have been explored across a wide range of engineering applications, with various forward modeling techniques and optimization algorithms tailored to specific use cases. 
However, most of these studies do not consider internal heat generation, which is a particularly important factor in electrical systems such as motors and power electronic devices. In addition, thermal conductivity is often the sole focus, whereas other important parameters, including density, specific heat capacity, and internal geometry, are frequently overlooked. This study addresses both challenges.

This paper is the second part of a two-part study. It presents a fast inverse heat conduction modeling approach for estimating uncertain thermal and geometric parameters in multi-layer composites with internal heat generation. Building on the forward model developed in Part I, this study explores its integration into various inverse modeling scenarios. 
A brief overview of the analytical forward thermal model is provided in Section~\ref{section: forward modeling}. Section~\ref{section: inverse modeling} details the construction of five-layer and ten-layer inverse models. The five-layer IHCM is used to study variations in material properties due to manufacturing tolerances, thermal aging, and degradation. The ten-layer IHCM addresses geometric changes such as global and local delamination. Corresponding finite element (FE) models generate reference temperature datasets and validate the inverse results. Specifically, 120 cases of unknown thermal conductivity and heat capacity of the epoxy layer, representing varying levels of thermal aging, are precisely predicted using the five-layer IHCM. Additionally, 60 cases of unknown position and thickness of global delamination, and 45 cases of local delamination with varying width and thickness ratios, are successfully inverted. Results reveal a correlation between delamination geometry and its thermal impact.
To evaluate the model's generality, nine thermal properties are selected as the potential unknowns, and ten test cases are demonstrated with different combinations of unknowns. These simulations assess the computational speed and convergence behavior of the inverse procedure.
Furthermore, a sensorless approach is proposed and tested across all cases, demonstrating strong potential for non-invasive thermal diagnostics in real-life applications.

\section{Forward modeling} \label{section: forward modeling}
A fast and accurate forward model is essential for the development of a reliable inverse model. In Part I of this research, analytical solutions for forward heat conduction in an arbitrary number of layers with diverse heat generations are derived and validated with a five-layer composite structure representative of a permanent magnet linear synchronous motor (PMLSM). This section provides an overview of those forward modeling solutions, including the treatment of various boundary conditions and heat generation conditions.

\subsection{Physical model} \label{section: physical model}
\begin{figure}
    \centering
    \includegraphics[width=0.6\textwidth]{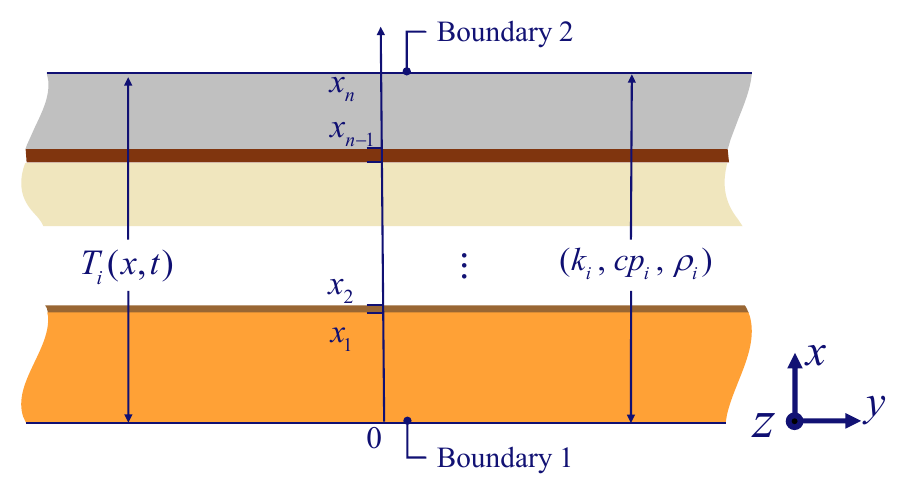}
    \caption{Generalized $n$-layer composite structure.}
    \label{fig:5 layer structure}
\end{figure}

A general $n$-layer composite structure with distinct materials of each layer is depicted in Fig.~\ref{fig:5 layer structure}. Perfect thermal contact is assumed at the interfaces, and internal heat generation may occur in one or more layers. 
The 1D heat diffusion equations along the x-axis are given by:
\begin{equation}
    \rho_i cp_i \frac{\partial T_i(x,t)}{\partial t}=  k_i\frac{\partial^2 T_i(x,t)}{\partial x^2}+\dot{q}_i(x,t),
\end{equation}
where $T_i(x,t)$ is the temperature distribution in the $i$-th layer, as a function of position $x$ and time $t$. The parameters $k_i$, $\rho_i$ and $cp_i$ denote the corresponding thermal conductivity, density, and specific heat capacity of the $i$-th layer. The term $q_i(x,t)$ represents the internal heat generation among the $i$-th layers, which may vary with both position and time. This partial differential equation (PDE) needs to be solved for each layer to obtain the full temperature distribution across the multi-layer composite, which was derived in detail in Part I of this study. 

Following Part I of this study, a five-layer composite part from a three-phase, six-stack PMLSM is used as the representative example. The thermophysical properties of the materials and geometry dimension of the layers are summarized in Table~\ref{tab:MateirialProperties}.
\begin{table}[H]
\begin{threeparttable}
    \centering
    \caption{Thermophysical properties and thickness of the layers}
    \begin{tabular}{l l c c c c}
        \hline
        \hline
        $i$ & Layer & $k_i$ & $cp_i$ & $\rho_i$ & $l_i$ \\
        \ &  &$\mathrm{[\frac{W}{m \cdot K}]}$ & $\mathrm{[\frac{J}{kg \cdot K}]}$ & $\mathrm{[\frac{kg}{m^3}]}$ & [$\mathrm{mm}$] \\
        \hline
         1 & Coil & 336 & 404 & 7734 & 0.78\\
         2 & Enamel & 0.4 & 1100 & 1300 & 0.02 \\
         3 & Epoxy & 1.3 & 800 & 2200 & 0.2 \\
         4 & Kapton & 0.4 & 1100 & 1300 & 0.025\\
         5 & Stainless steel & 16.3 &  500 & 8000 & 0.25\\ 
         \hline
         \hline
    \end{tabular}
    \label{tab:MateirialProperties}
    \begin{tablenotes}
    \footnotesize
    \item[*] Coil layer consists of copper wires coated with polyimide enamel insulation.
    \end{tablenotes}
    \end{threeparttable}
    \end{table} 
Heat is generated in the coil layer through ohmic losses from a 25 A current excitation. For this five-layer structure, a symmetrical boundary condition (Neumann type) is applied at boundary 1, while a convective cooling boundary condition
(Robin type) is applied at boundary 2. At each interface between layers, the continuous boundary conditions should be satisfied to ensure that both temperature and heat flux remain continuous across the boundaries. These boundary conditions are expressed as:
\begin{equation}
    \begin{split}
        \frac{\partial T_1}{\partial x}(x,t)|_{x=0}&=0,\\
        T_i(x,t)|_{x=x_i}&=T_{i+1}(x,t)|_{x=x_i}, \   (i=1,...,n-1),\\
        k_i\frac{\partial T_i(x,t)}{\partial x}\Big|_{x=x_i}&=k_{i+1}\frac{\partial T_{i+1}(x,t)}{\partial x}\Big|_{x=x_i}, \  i=(1,...,n-1), \\
        k_n \frac{\partial T_n(x,t)}{\partial x}\Big|_{x=x_n}&=-h(T_n(x,t)|_{x=x_n}-T_{\text{co}}),
    \end{split}
    \label{eq:BCs}
\end{equation}
where $T_{\text{co}}$ denotes the coolant temperature, and $h=1050\ \mathrm{W/(m^2\cdot K)}$ represents the heat transfer coefficient at the coolant-stainless steel interface, as estimated in~\citep{fuAnisotropic3DThermal2024}. 
The initial temperature is assumed to be uniform across the entire composite, as the system typically begins from a stable ambient temperature. This is given as $T_i(x,0)=T_{0}$, where $T_{0}$ represents the room temperature. In this study, both $T_{\text{co}}$ and $T_0$ are set to $20 \mathrm{^\circ C}$.

\subsection{Solution of the forward model}
\subsubsection{SOV-OE approach}
A combined modeling approach based on the separation of variables (SOV) method~\citep{davidw.hahnHeatConduction2012,zhouTheoreticalSolutionTransient2017,demonteTransientHeatConduction2000,polyaninSeparationVariablesExact2021} and the orthogonal expansion (OE) technique~\citep{demonteAnalyticApproachUnsteady2002,braunDifferentialEquationsTheir1993,demonte2003unsteady} is presented in Part I to account for the case with constant heat generation, where $\dot{q}_1(x,t)=\dot{q}_g$. In this SOV-OE approach, the temperature solution of each layer is first separated as the product of a spatial component $\mathcal{X}(x)$ and a temporal component $\mathcal{T}(t)$:
\begin{equation}
    T_i(x,t)=\mathcal{X}_i(x)\mathcal{T}_i(t),
\end{equation}
where $\mathcal{X}(x)$ and $\mathcal{T}(t)$ are solved separately in the forms of the Helmholtz equation and the exponential equation, respectively:
\begin{equation}
\begin{split}
    \mathcal{X}_i(x)&=a_i sin(\lambda x/\sqrt{\alpha_i})+b_i cos(\lambda x/\sqrt{\alpha_i}), \\
    \mathcal{T}_i(t)&=c_i e^{-\lambda^2 t}.
\end{split}    
\label{eq:SOV method solutions}
\end{equation}
These components contain unknown eigenvalues $\lambda$ and coefficients $a_i$, $b_i$, $c_i$, which need to be determined by applying the boundary conditions. This forms a regular Sturm-Liouville problem, which allows the spatial component $\mathcal{X}_i(x)$ to be expanded as a convergent series of eigenfunctions $\mathcal{X}_{i,m}(x)$~\citep{braunDifferentialEquationsTheir1993}. The general solution for the temperature distribution $T_i(x,t)$ in each layer is then expressed as: 
\begin{equation}
\begin{split}
    T_i(x,t) &= \sum_{m=1}^\infty C_m e^{-\lambda_m^2 t} \mathcal{X}_{i,m}(x) +D_i x+E_i-\dot{q}_ix^2/2k_i,
    \label{eq:T_sp}
\end{split}
\end{equation}
where $D_i$ and $E_i$ are integration constants from the boundary conditions. The coefficient $C_m$, known as the Fourier coefficient~\citep{davidw.hahnHeatConduction2012}, is obtained by applying the initial condition using the orthogonal expansion technique\citep{davidw.hahnHeatConduction2012,demonteAnalyticApproachUnsteady2002,demonteTransientHeatConduction2000}:
\begin{equation}
 C_m=\frac{1}{N_m}\sum^n_{i=1}  \int^{x_i}_{x_{i-1}} \frac{k_i}{\alpha_i} F_i(x)\mathcal{X}_{i,m}(x) dx,
 \label{eq:fourier coefficient}
\end{equation}
where $N_m$ is the norm of the eigenfunctions, defined as:
\begin{equation}
      N_m=  \sum^n_{i=1} \int^{x_i}_{x_{i-1}} \frac{k_i}{\alpha_i}  (\mathcal{X}_{i,m}(x))^2 dx, 
\end{equation}
and $F_i(x)$ is given by:
\begin{equation}
    F_i(x) = T_0 - D_i x - E_i + \dot{q}_ix^2/2k_i.
\end{equation}
This framework provides an exact closed-form analytical solution for the temperature distribution in an $
n$-th layer composite subjected to constant internal heat generation. 

\subsubsection{Green's function-based approach}
For problems involving transient heat generation and non-homogeneous conditions, the SOV-OE approach faces limitations due to its requirement for recomputing coefficients at each time step. To overcome these challenges, a Green’s function-based (GF-based) approach is introduced and thoroughly explained in Part I of this study. This method leverages the solution structure derived from the SOV-OE formulation and enables a more flexible and computationally efficient solution. The GF-based approach expresses the temperature solution as a superposition of individual contributions from the initial condition (IC), heat generation (HG), and boundary conditions (BCs). The overall temperature solution can be written in the general form~\citep{davidw.hahnHeatConduction2012,ozisik1993heat,cole2010heatgreensfunction}:
\begin{equation}
    \begin{split}
        T_i(x,t) &= = \theta_{\text{IC}}(x,0) + \theta_{\text{HG}}(x,t) + \theta_{\text{BCs}}(x,t) \\
        &= \sum_{j=1}^n  \int_{x_{j-1}}^{x_j} G_{i,j} (x,t|x',\tau)|_{\tau=0} T_{\text{IC}}(x')dx' \\
        &+ \sum_{j=1}^n \int_{\tau=0}^t \int_{x_{j-1}}^{x_j} G_{i,j}(x,t|x',\tau) \frac{\alpha_j}{k_j} \dot{q}_j(x',\tau)dx' d\tau \\
        &+ \left[\int_{\tau=0}^t G_{i,j}(x,t|x',\tau)|_{x'=0} \frac{\alpha_1}{k_1} f_1(x',\tau)  d\tau + \int_{\tau=0}^t G_{i,j}(x,t|x',\tau)|_{x'=x_n} \frac{\alpha_n}{k_n} f_n(x',\tau)  d\tau \right],
    \end{split}
    \label{eq:1D_GF_solution}
\end{equation}
where the temperature contributions $\theta_{\text{IC}}$, $\theta_{\text{HG}}$, and $\theta_{\text{BCs}}$ are expressed as the convolution of the Green's function with the corresponding sources or conditions. The term $T_{\text{IC}}(x')$ represents the initial temperature at position $x'$, and $f_i(x',\tau)$ denotes the time-dependent non-homogeneous boundary term. The general form of the Green's function is given by:
\begin{equation}
 G_{i,j}(x,t|x',\tau)= \sum_{m=1}^\infty  \frac{1}{N_m} \frac{k_j}{\alpha_j} e^{-\lambda_m^2 (t-\tau)} \mathcal{X}_{i,m}(x) \mathcal{X}_{j,m}(x').
\label{eq:general GF}
\end{equation}
This approach provides a powerful tool to accurately calculate the temperature distribution while effectively accounting for non-homogeneities introduced by boundary conditions and transient heat sources.

In summary, the SOV-OE and GF-based methods provide fast and accurate forward solution that are well-suited for integration into inverse heat conduction models (IHCMs). Detailed derivation, implementation, and validation of these forward models are presented in Part I of this research and will not be repeated here. In the next section, these forward models are employed to construct IHCMs tailored to various inverse scenarios.
\nomenclature{$h$}{Heat transfer coefficient [$\mathrm{W/(m^2 \cdot K)}$]}
\nomenclature{$T_{\text{co}}$}{Coolant temperature [$\mathrm{K}$]}
\nomenclature{$T_{0}$}{Initial/room temperature [$\mathrm{K}$]}
\nomenclature{$R_{0}$}{Resistance of the coil under reference temperature [$\Omega$]}
\nomenclature{$k_i$}{Thermal conductivity of the $i$th layer[$\mathrm{W/(m \cdot K)}$]}
\nomenclature{$k_{\text{epo}}$}{Thermal conductivity of the epoxy [$\mathrm{W/(m \cdot K)}$]}
\nomenclature{$\rho_i$}{Mass density of the $i$th layer [$\mathrm{kg/m^3}$]}
\nomenclature{$T_i(x,t)$}{Temperature of the $i$th layer [$\mathrm{K}$]}
\nomenclature{$T_{\text{ave}}$}{Average temperature of the coil layer [$\mathrm{K}$]}
\nomenclature{$T_{\text{ref}}$}{Reference temperature dataset [$\mathrm{K}$]}
\nomenclature{$T^*$}{Iterative temperature dataset [$\mathrm{K}$]}
\nomenclature{$cp_i$}{Specific heat capacity of the $i$th layer [$\mathrm{J/(kg \cdot K)}$]}
\nomenclature{$cp_{\text{epo}}$}{Specific heat capacity of the epoxy [$\mathrm{J/(kg \cdot K)}$]}
\nomenclature{$x, y, z$}{Spatial coordinate [$\mathrm{mm}$]}
\nomenclature{$n$}{Number of layers}
\nomenclature{$l_i$}{Thickness of the $i$th layer [$\mathrm{mm}$]}
\nomenclature{$t$}{Time [$\mathrm{s}$]}
\nomenclature{$\dot{q}_i$}{Volumetric heat generation rate of the $i$-th layer [$\mathrm{W/m^3}$]}
\nomenclature{$h$}{Heat transfer coefficient [$\mathrm{W/(m^2 \cdot K)}$]}
\nomenclature{$G$}{Green's function}
\nomenclature{$f$}{Boundary conditions}
\nomenclature{$\tau$}{Time of heat source}
\nomenclature{$\theta$}{Temperature contribution\\
-$\theta_{IC}$ from initial condition\\
-$\theta_{HG}$ from heat generation\\
-$\theta_{BCs}$ from boundary conditions\\}
\nomenclature{$a, b, c$}{Coefficients of integration}
\nomenclature{$D, E$}{Integration constants}
\nomenclature{$C$}{Fourier coefficients}
\nomenclature{$F$}{Initial conditions}
\nomenclature{$N$}{Norm}
\nomenclature{$\mathcal{X}$}{Spatial component of temperature}
\nomenclature{$\mathcal{T}$}{Temporal component of temperature}
\nomenclature{$U$}{Terminal voltage of the coil winding [$\mathrm{V}$]}
\nomenclature{$I$}{Current excitation of the coil winding [$\mathrm{A}$]}
\nomenclature{$\beta$}{Temperature coefficient of copper resistivity[$\mathrm{K^{-1}}$]}
\nomenclature{$w_{\text{dela}}$}{Width of the local delamination [$\mathrm{mm}$]}
\nomenclature{$w_{\text{coil}}$}{Width of the single side of the coil [$\mathrm{mm}$]}
\nomenclature{$wr$}{Width ratio of the local delamination to the entire width of a single side for a coil winding}
\nomenclature{IHCM}{Inverse heat conduction model}
\nomenclature{IHCP}{Inverse heat conduction problem}

\section{Inverse modeling} \label{section: inverse modeling}

The inverse heat conduction problem is commonly concerned to identify unknown quantities in the mathematical formulation of heat conduction using measurements of system response~\citep{colacoInverseOptimizationProblems2006}, for instance temperature data. These unknowns may include thermal conductivity, heat flux, heat transfer coefficient, or internal geometric features, etc.

This section presents the implementation of inverse procedures and validates the results through FE simulations. Two composite configurations from a permanent magnet linear synchronous motor (PMLSM), namely a five-layer and a ten-layer structure, are used to demonstrate the inverse modeling approach. The five-layer model is used for symmetrical scenarios, whereas the ten-layer model is applied to asymmetrical ones. Various inverse scenarios are investigated to address practical challenges, such as detecting changes in thermal properties due to thermal aging, and identifying the location and size of the internal delamination. In addition, a non-invasive sensorless measuring strategy is introduced and tested within the inverse modeling framework. The feasibility, sensitivity, and computational efficiency of the proposed IHCMs are evaluated across various test cases.

\subsection{Inverse model}

\begin{figure}[h!]
    \centering
    \includegraphics[width=0.5\textwidth]{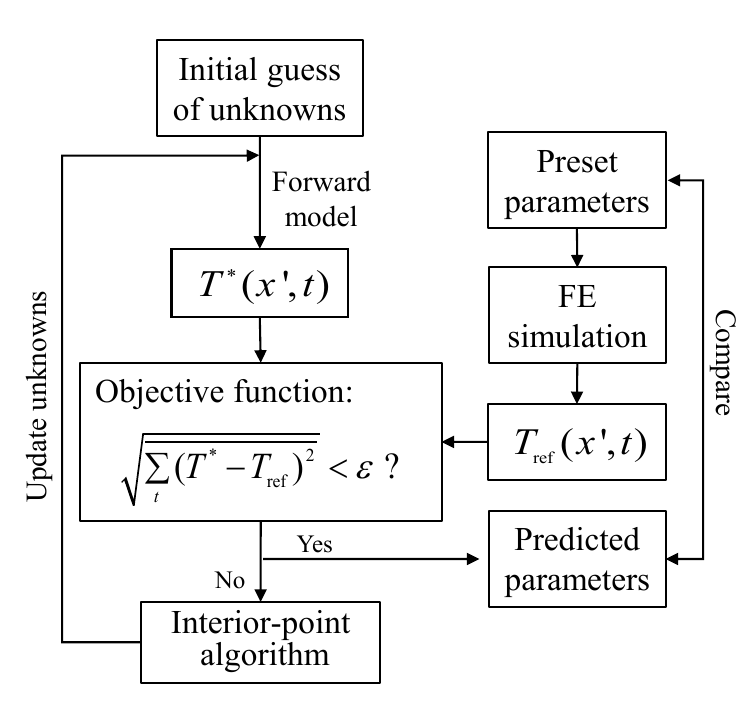}
    \caption{Framework of the inverse modeling procedure based on temperature datasets from FE simulations.}
    \label{fig:inverse procedures}
\end{figure}

The forward models discussed earlier serve as the foundation for the inverse modeling framework. They are integrated into an iterative process to estimate unknown parameters by matching calculated and reference temperature profiles. The inverse procedure involves the following steps, as illustrated in Fig.~\ref{fig:inverse procedures}:

1. \textbf{Initialization:} Initial guesses for the unknown parameters are provided within reasonable bounds, which may be based on data sheets or expert knowledge.

2. \textbf{Forward Simulation:} The forward model is executed using the initial guesses to compute a temperature profile $T^*(x',t)$.

3. \textbf{Objective Function Evaluation:} The discrepancy between the computed profile $T^*(x',t)$ and the reference data $T_{\text{ref}}(x',t)$ is quantified using the sum of squared mean errors (SSME) as the objective function.

4. \textbf{Parameter Optimization:} The interior point optimization algorithm is employed to update the unknown parameters. If the defined tolerance $\varepsilon$ is not met, steps 2 and 3 are repeated until convergence.

5. \textbf{Validation:} Once the objective function meets the tolerance, the predicted parameters are compared against the known preset values for validation.

The reference temperature profiles $T_{\text{ref}}(x',t)$ are generated using FE simulations with predefined values for the unknown parameters. The point of temperature measurement, $x'$, is taken at the enamel–epoxy (E–E) interface in this study. Optimization is carried out using the interior point algorithm via MATLAB’s `fmincon' function, which requires lower and upper bounds for the unknown parameters. These bounds are set to 50$\%$ and 200$\%$ of the nominal values from the material datasheet. The maximum number of iterations is set to 200, and the convergence tolerance is defined as $10^{-3}$, which terminates the optimization either upon reaching the maximum iteration count or when constraint violations fall below the threshold.

\begin{figure}[h!]
    \centering
    \includegraphics[width=0.48\textwidth]{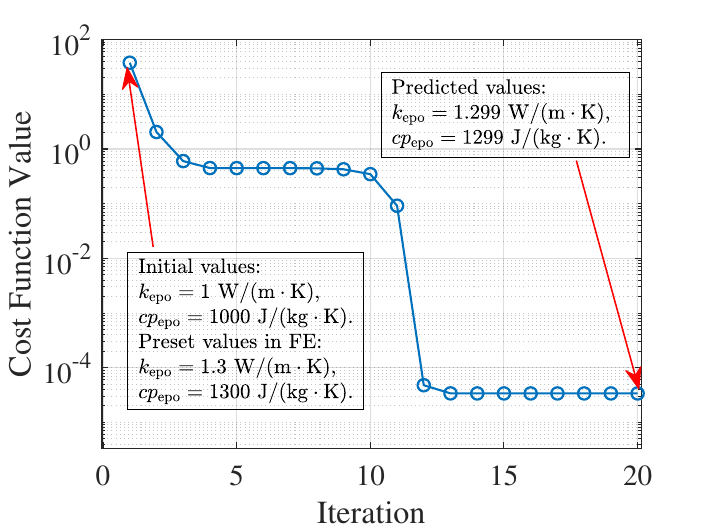}
    \caption{Convergence behavior of the IHCM for estimating two unknowns $k_{\text{epo}}$ and $cp_{\text{epo}}$, using an initial guess of $k_{\text{epo}} = 1 \  \mathrm{W/(m\cdot K)}$ and $cp_{\text{epo}} = 1000 \ \mathrm{J/(kg\cdot K)}$. Reference temperature dataset are generated from a 1D FE model, at the E-E interface, with preset values of $k_{\text{epo}} = 1.3 \  \mathrm{W/(m\cdot K)}$ and $cp_{\text{epo}} = 800 \ \mathrm{J/(kg\cdot K)}$.}
    \label{fig:convergence_objective function}
\end{figure}

To illustrate the implementation of this IHCM, a simple case is presented where the thermal conductivity $k_{\text{epo}}$ and specific heat capacity $cp_{\text{epo}}$ of the epoxy layer are treated as unknowns. The reference temperature dataset is obtained from a 1D FE model using preset values: $k_{\text{epo}} = 1.3 \  \mathrm{W/(m\cdot K)}$, and $cp_{\text{epo}} = 800 \ \mathrm{J/(kg\cdot K)}$. The initial guesses are set to $k_{\text{epo}} = 1 \  \mathrm{W/(m\cdot K)}$ and $cp_{\text{epo}} = 1000 \ \mathrm{J/(kg\cdot K)}$. After several iterations, the IHCM converges, and the predicted values closely match the reference values, as the convergence behavior of the objective function shown in Fig.~\ref{fig:convergence_objective function}. This example result validates the effectiveness of the proposed IHCM, and more complex scenarios will be explored in the following sections using the same validation framework.

\subsection{Case 1: Unknown material properties} \label{section: thermal aging}

In devices composed of multi-layer structures operating under transient thermal conditions, accurate material thermal properties are essential for reliable thermal performance evaluation. However, deviations from the nominal properties may arise due to factors such as manufacturing tolerances, material impurities, or long-term operational effects including thermal aging and material degradation. These changes can lead to biased evaluation on temperature distribution if standard datasheet values are used. In such cases, an IHCM offers a promising way to calibrate or monitor actual material properties based on minimized internal temperature measurements.

This case study investigates the five-layer composite structure introduced in Section~\ref{section: physical model}, with material properties listed in Table~\ref{tab:MateirialProperties}. The epoxy layer is assumed to experience thermal stress over long-term operation, resulting in reduced thermal conductivity $k_{\text{epo}}$ and specific heat capacity $cp_{\text{epo}}$. The initial values are set to $k_{\text{epo}}=1.9 \ \mathrm{W/(m \cdot K)}$ and $cp_{\text{epo}}=1500 \ \mathrm{J/(kg \cdot K)}$. To represent different degradation levels, $k_{\text{epo}}$ is varied from 0.8 to 1.9 and $cp_{\text{epo}}$ from 600 to 1500, yielding 120 distinct combinations.

Corresponding to each of these 120 combinations, reference temperature datasets are generated at the E-E interface using the 1D FE model. These reference datasets serve as the input for the objective function of the IHCM. And the IHCM is then used to estimate the values of $k_{\text{epo}}$ and $cp_{\text{epo}}$ based on each dataset, resulting in 120 predicted parameter pairs.

\begin{figure*}
    \centering
    \includegraphics[width=1\linewidth]{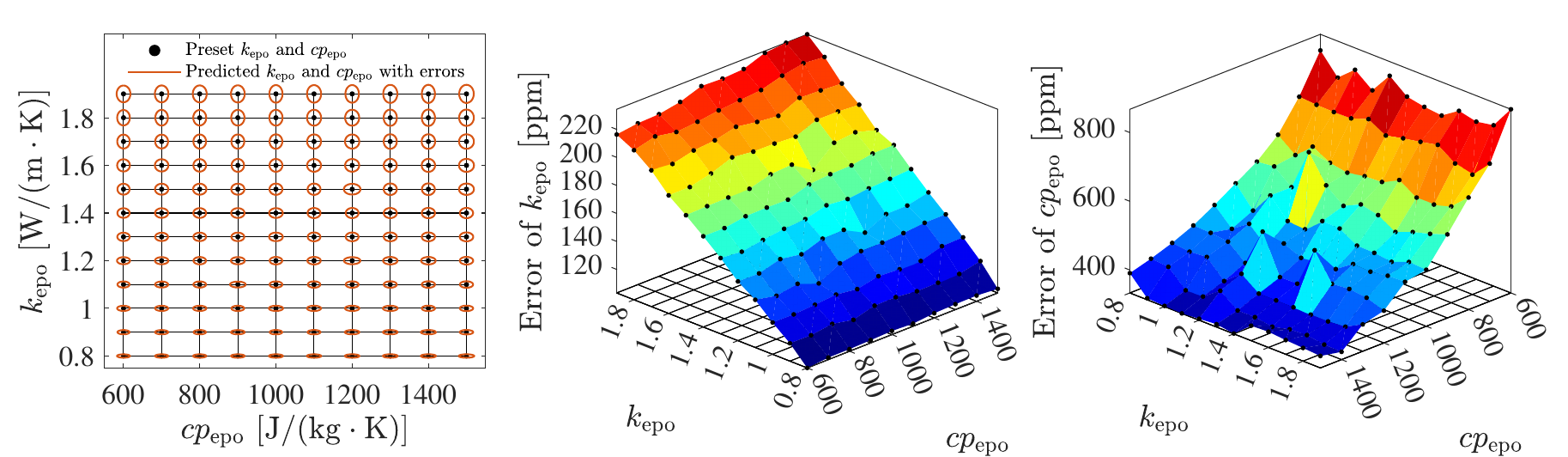}
    \caption{Inverse results of the five-layer IHCM under a thermal aging scenario, with $k_{\text{epo}}$ and $cp_{\text{epo}}$ as the unknowns. Reference temperature datasets are generated from 1D FE simulations at the E-E interface using 120 combinations of preset values: $k_{\text{epo}}$ (0.8$\sim$1.9 $\mathrm{W/(m \cdot K)}$) and $cp_{\text{epo}}$ (600$\sim$1500 $\mathrm{J/(kg \cdot K)}$), representing varying levels of thermal aging.}
    \label{fig:1d aging}
\end{figure*}

The inverse results are visualized in a checkerboard plot in Fig.~\ref{fig:1d aging}, where the relative errors are qualified by the short- and long-axis lengths of the ellipses. Accompanying error maps for $k_{\text{epo}}$ and $cp_{\text{epo}}$ are also presented. Where “ppm” (parts per million) denotes a high level of agreement between the predicted and preset values in the FE models. 
A trend is observed where larger error occurs for combinations with higher $k_{\text{epo}}$ and lower $cp_{\text{epo}}$. This can be attributed to the systematic differences between the analytical forward model and the FE model. In cases with larger $k_{\text{epo}}$, the thermal resistance of the epoxy layer becomes less significant, making its influence on the overall temperature response weaker. As a result, the inverse predictions for such parameters are more prone to error. A similar principle applies to scenarios with lower $cp_{\text{epo}}$. These findings highlight that inverse predictions in layered structures are generally more accurate for parameters that have a more pronounced influence on the temperature distribution. 

\begin{figure*}
    \centering
    \includegraphics[width=1\linewidth]{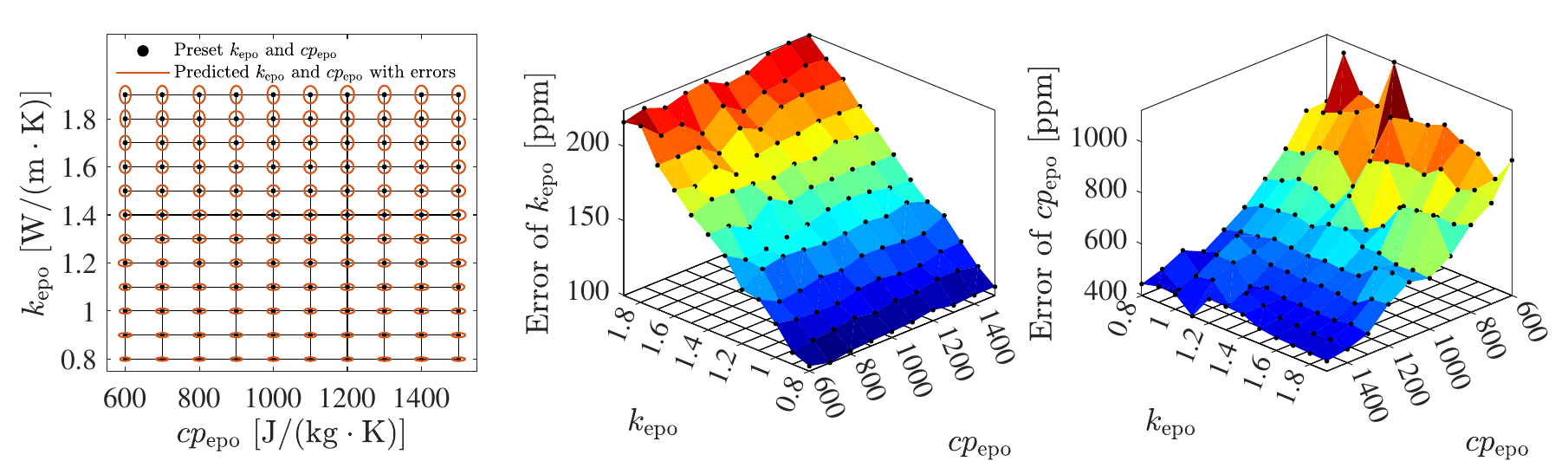}
    \caption{Inverse results of the five-layer IHCM under a thermal aging scenario, with $k_{\text{epo}}$ and $cp_{\text{epo}}$ as the unknowns. Reference temperature datasets are generated from 2D FE simulations at the E-E interface using 120 combinations of preset values: $k_{\text{epo}}$ (0.8$\sim$1.9 $\mathrm{W/(m \cdot K)}$) and $cp_{\text{epo}}$ (600$\sim$1500 $\mathrm{J/(kg \cdot K)}$), representing varying levels of thermal aging.}
    \label{fig:2d aging interface T no bobbin}
\end{figure*}

\begin{figure*}[h!]
    \centering
    \includegraphics[width=0.48\textwidth]{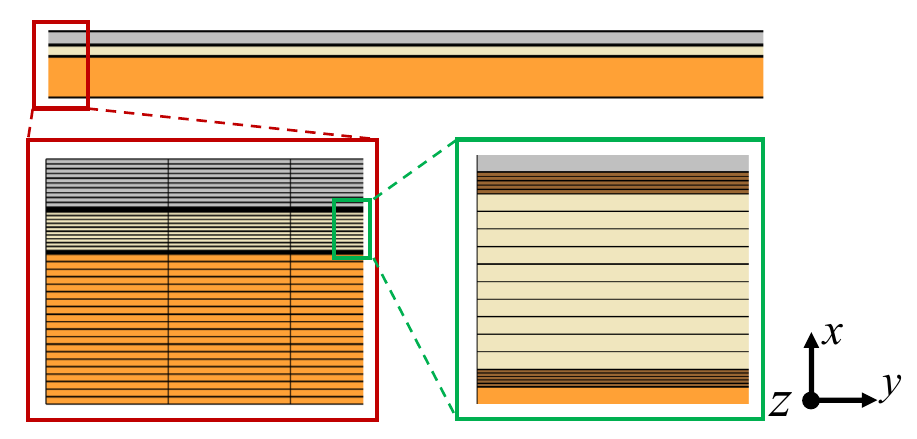}
    \caption{2D FE model with mesh configuration.}
    \label{fig:2D griding}
\end{figure*}

To further evaluate the robustness and compatibility of the IHCM, a 2D FE model is employed to generate additional reference datasets using the same preset values. 
In the 2D FE models, the temperature at the geometric center of the E-E interface is taken as the reference temperature. Following the same procedure, 120 groups of reference temperature datasets are generated from the 2D FE simulations, corresponding to the same combinations of $k_{\text{epo}}$ and $cp_{\text{epo}}$ used in the 1D case. These datasets are then used in the IHCM to estimate the unknown thermal properties. The inverse results are compared with the preset values and illustrated in Fig.~\ref{fig:2d aging interface T no bobbin}.
And the corresponding mesh setting is shown in Fig.~\ref{fig:2D griding}. 

The results show that errors from both the 1D and 2D FE-based inverse estimations are comparably small. This consistency arises because the FE simulations assume uniform heat generation and ideal thermal insulation boundaries in 2D FE model, which limit heat transfer along transversal ($y$) direction. 
However, in practical scenarios where lateral or radial heat transfer cannot be neglected, an alternative modeling strategy such as constructing a thermal equivalent circuit may be applied to account for multidimensional heat flow. Such approach helps mitigate the influence on inverse accuracy.

In some applications, simulation speed is a higher priority than accuracy, particularly for real-time condition monitoring systems that require rapid and continuous evaluation of thermal performance. Computational efficiency is also essential for implementation on portable devices with limited processing capabilities.
A comparison of simulation time for the analytical forward model, the 1D and 2D FE simulations, and the inverse modeling process is summarized in Table~\ref{tab:simulation time}.
\begin{table}[H]
    \begin{threeparttable}
       \centering
    \caption{Simulation time of models, with 7980x 64-cores CPU}
    \begin{tabular}{L C c c c c}
        \hline
        \hline
         & Numbers of  & Proposed & 1D & 2D\\
         &  simulations & model & FEM & FEM  \\
          \hline
        Forward & 1  &0.013 s & 2.5 s & 4 s \\
         modeling & 120  & 0.80 s & 179 s & 473 s\\
         \hline
         *Inverse & 1  & 0.67 s & - & -  \\
         modeling & 120  &  81 s & - & -  \\
         \hline
         \hline
    \end{tabular}
    \label{tab:simulation time}
    \begin{tablenotes}
    \footnotesize
    \item[*] This inverse model estimates two unknowns: $k_{\text{epo}}$ and $cp_{\text{epo}}$, with the number of layers $m=3$.
    \end{tablenotes}
    \end{threeparttable}
    \end{table}
Under identical spatial and temporal discretization, the analytical model achieves computation time approximately 30 and 60 times faster than the 1D and 2D FE models. Given that the inverse process involves multiple iterations of the forward model, this efficiency offers significant advantages. The total computation time of the inverse process depends on several factors, including the numbers of unknowns, the convergence tolerance, and the number of eigenvalues used. In the current case with two unknowns $k_{\text{epo}}$ and $cp_{\text{epo}}$ and three eigenvalues, a single inverse simulation takes approximately 0.67 seconds. The most computationally intensive part is determining the eigenvalues $\lambda$ at each iteration, where the simulation time scales nearly quadratically with the number of $\lambda$. This forms a tradeoff between computational cost and accuracy, while more efficient eigenvalue search algorithms could further accelerate the inverse process. 

In summary, this case study demonstrates the IHCM's capability to estimate thermally degraded material properties efficiently. While $k_{\text{epo}}$ and $cp_{\text{epo}}$ were selected here as representative unknowns, any parameter appearing in the diffusion equation may be designated for inversion. The proposed approach is not limited to the five-layer PMLSM example, but is applicable to a broad range of multi-layer structures with internal heat generation and arbitrary layer configurations and boundary conditions.

\subsection{Case 2: Unknown geometric dimensions} \label{section: delamination}

During prolonged operation, machines experience not only variations in material properties, but also changes in geometry. Machines that produce motion output, such as rotatory motors, linear motors, and planar motor~\citep{encicaElectromagneticThermalDesign2008,jansenOverviewAnalyticalModels2014,roversAnalysisMethodDynamic2012,curtiOverviewAnalyticalMethods2015,curtiGeneralFormulationMagnetostatic2018}, may undergo chafing at the layer interfaces due to repeated acceleration and deceleration. Meanwhile, thermal expansion occurs in multi-layer composites subjected to thermal stress, where materials with different thermal expansion coefficients create shear force at the interfaces. This has been observed in applications such as power cables, power electronic devices, and semiconductor manufacturing. Consequently, the mechanical motion and thermal expansion in these systems can cause shearing and chafing at the interfaces, eventually leading to geometric deformations such as delamination or other structural variations~\citep{yangInSituDelamination2023,congDetectionPrintedCircuit2022,gillespieCompositeLaminateDelamination2020,huaiDetectingLocalDelamination2021}. In addition, high-temperature or high-voltage environments may cause polymer materials to degrade and release gases within the layers, resulting in the formation of voids or gas pockets.

\begin{figure*}
    \centering
    \includegraphics[width=0.8\linewidth]{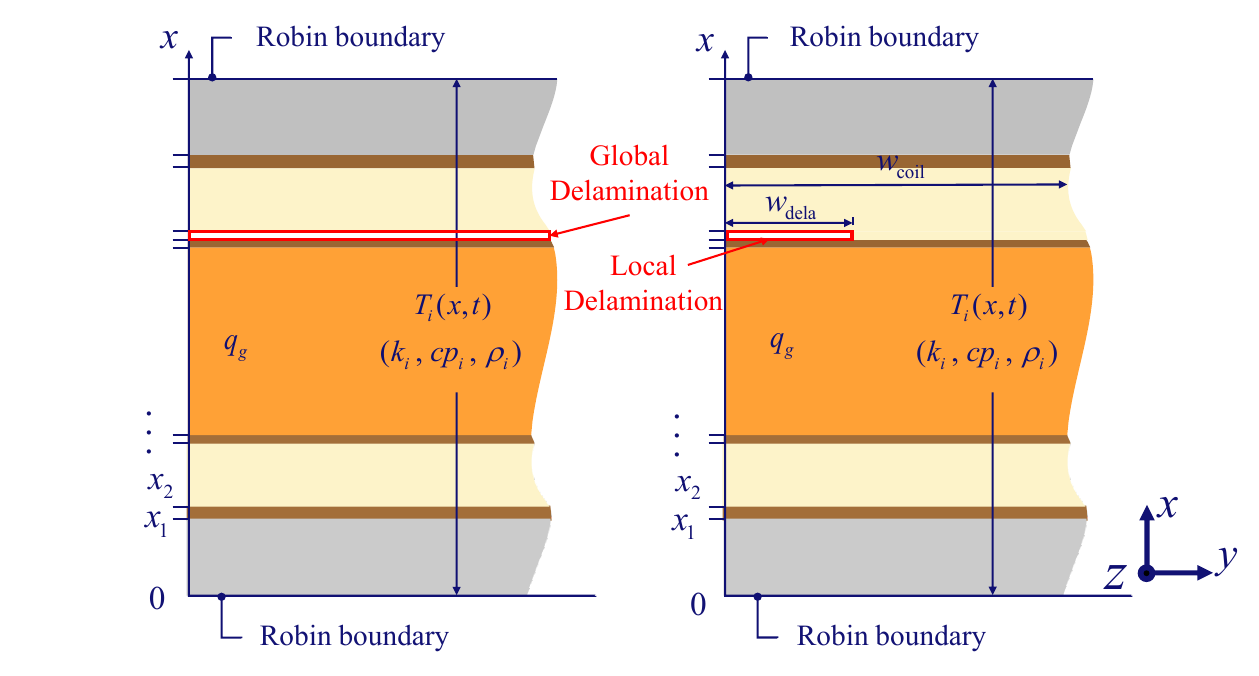}
    \caption{Schematic illustration of the global (left) and local (right) delamination at the E-E interface using asymmetrical models with the number of layers $n=10$.}
    \label{fig:delamination sketch}
\end{figure*}

In this case study, both global and local delamination scenarios are modeled using 1D and 2D FE simulations to generate the reference temperature datasets. An additional air layer is introduced at specific interfaces to mimic delamination, with its location and thickness treated as unknowns to be identified by the IHCM. 
Since delamination does not always occur symmetrically on both sides of the coil, the assumption of symmetry about the coil center is not applicable. Therefore, both the FE and analytical models are adjusted to represent an asymmetrical configuration with ten layers ($n=10$) in total. Cooling (Robin-type) boundary conditions are applied to the top and bottom surfaces, as illustrated in Fig.~\ref{fig:delamination sketch}.
Where the first boundary condition in Eq.~\ref{eq:BCs} should be replaced by:
\begin{equation}
        k_1 \frac{\partial T_1(x,t)}{\partial x}\Big|_{x=0}=h(T_1(x,t)|_{x=x_1}-T_{\text{co}}).
    \label{eq:BCs_modified}
\end{equation}

\subsubsection{Global delamination}
\begin{figure*}
    \centering
    \includegraphics[width=0.8\linewidth]{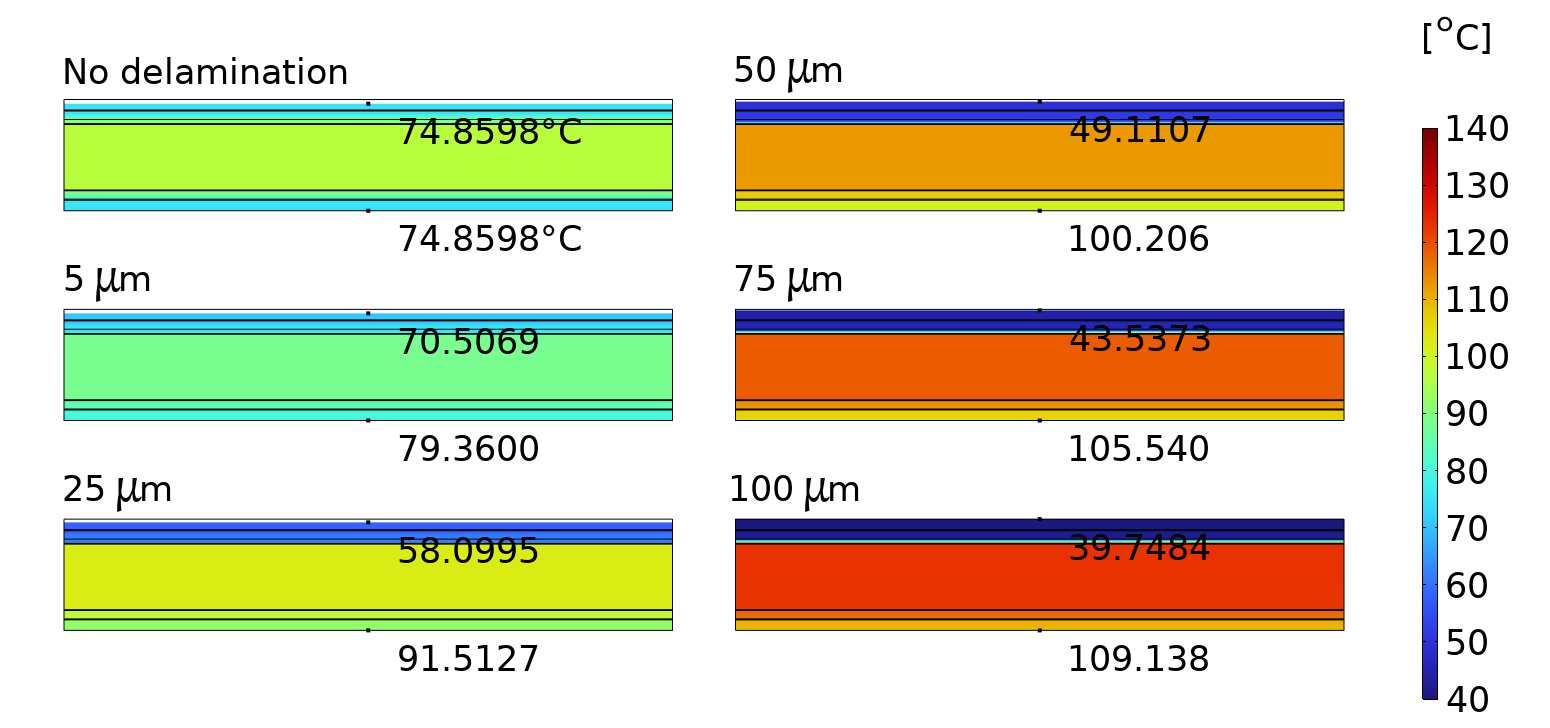}
    \caption{Temperature distribution of the asymmetrical 2D FE models at t = 25 s, for delamination thickness of 0, 5, 25, 50, 75, and 100 $\mathrm{\mu m}$.} 
    \label{fig:temperature distribution with diff dela}
\end{figure*}

Global delamination refers to a complete layer, filled with air across the full width of the interface. Variations in the location and thickness of such a delaminated layer result in distinct temperature distributions across the composite. To demonstrate this, temperature fields are simulated in 2D FE models, where an air layer is inserted at the E-E interface. The temperature distributions for various delamination thicknesses are shown in Fig.~\ref{fig:temperature distribution with diff dela}.
The results reveal that the presence of delamination layer increases the temperature of the layers below it. This occurs because the delaminated air layer impedes heat flow toward the upper cooling surface, causing heat to accumulate below. In contrast, the temperature above the delamination layer becomes lower. As the delamination thickness increases, this thermal separation phenomenon becomes more pronounced, leading to greater temperature differences between the upper and lower parts of the structure. These findings suggest that delamination not only creates localized thermal stress but may also increase thermal loading on healthy layers in the other side, which can accelerate the formation of new defects. This highlights the importance of early delamination detection to prevent cascading thermal failures. Additionally, the location of the delamination determines which layers are exposed to elevated stress levels.

\begin{figure}
    \centering
    \includegraphics[width=0.6\textwidth]{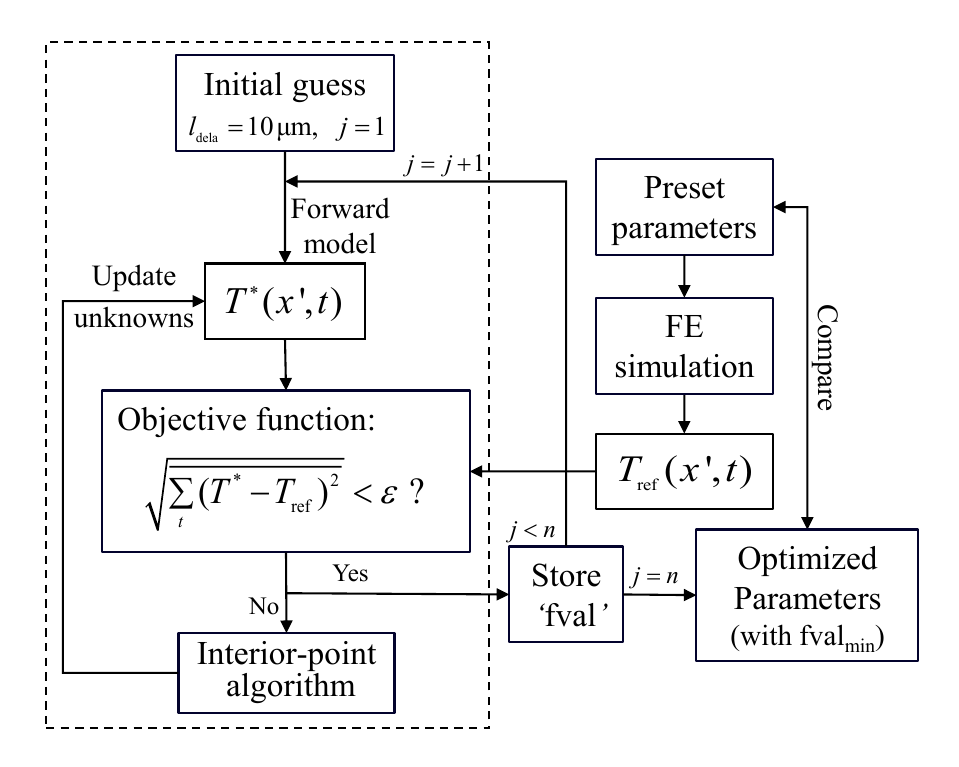}
    \caption{Inverse modeling procedures for detecting delamination.} 
    \label{fig:inverse procedures with diff dela}
\end{figure}

To predict both the thickness and location of the delamination, the inverse model needs to incorporate an additional loop, as illustrated in Fig.~\ref{fig:inverse procedures with diff dela}. Where the index $j$ represents the assumed delamination location, and the value of the objective function is denoted by `\text{fval}'. The delamination thickness $l_{\text{dela}}$ is treated as the unknown to be optimized, while the location is inferred by selecting the $j$ that minimizes the objective function.

\begin{figure}
    \centering
    \includegraphics[width=0.48\textwidth]{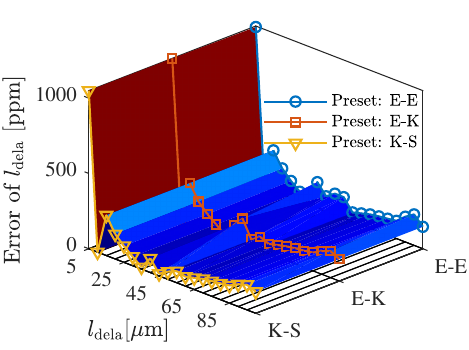}
    \caption{Inverse results of the ten-layer IHCM under a global delamination scenario, with delamination location and $l_{\text{dela}}$ as the unknowns. Reference temperature datasets are generated from 2D FE simulations using 60 combinations of preset values: three delamination locations (E-E, E-K, and K-S interfaces) and $l_{\text{dela}}$ ($5\sim100\ \mathrm{\mu m}$), representing different global delamination conditions.} 
    \label{fig:inverse results dela global}
\end{figure}

In this case study, delamination is modeled at three possible interfaces: enamel-epoxy (E-E), epoxy-Kapton (E-K), and Kapton-stainless steel (K-S). For each location, twenty thicknesses ranging from 5 to 100 $\mathrm{\mu m}$ are tested, resulting in 60 reference temperature datasets. These datasets are generated from both 1D and 2D FE simulations and applied to the inverse model to predict the corresponding delamination parameters. 
The inverse results, shown in Fig.~\ref{fig:inverse results dela global}, indicate that the predicted locations precisely match the preset delamination locations. The errors in the predicted delamination thicknesses remain below 1000 ppm. The largest error is observed when the delamination thickness is $5 \mathrm{\mu m}$, likely due to systematic discrepancies between the analytical and FE models. This effect becomes more pronounced when thinner layers are involved.
These results confirm that the inverse modeling approach can effectively detect both the occurrence and the thickness of delamination, as well as identify the interface and the side of the coil where it occurs. 

\subsubsection{Local delamination}
\begin{figure}
    \centering
    \includegraphics[width=0.48\textwidth]{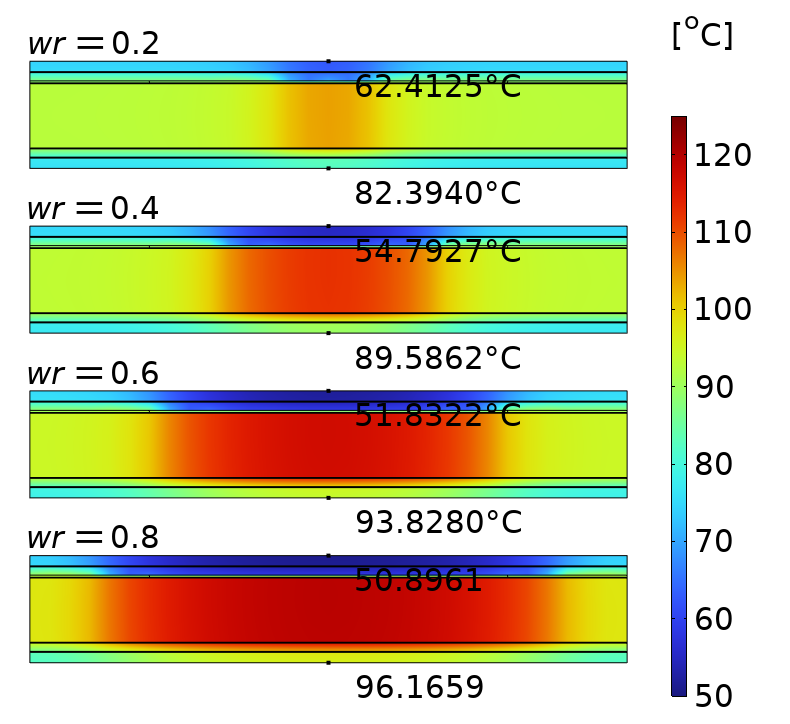}
    \caption{Temperature distribution of the asymmetrical 2D FE models at t = 25 s, and $l_{\text{dela}} = 50 \ \mathrm{\mu m}$, for width ratio $wr$ of 0.2, 0.4, 0.6, 0.8 $\mathrm{\mu m}$.} 
    \label{fig:temperature distribution with diff wr}
\end{figure}

Global delamination is common in compact structures with relatively uniform stress distribution, while local delamination is also frequently observed. In such cases, shear stress is unevenly distributed across the layers, causing partial separations at specific spots. 
To investigate how the size of local delamination influences the inverse results, localized air gaps are modeled in 2D FE simulations at the E-E interface on the upper side of the coil. The local delaminations vary in both thickness (5, 25, 50, 75, 100 $\mathrm{\mu m}$) and width ratio (0.1$\sim$0.9), defined as $wr = w_{\text{dela}}/w_{\text{coil}}$, where $w_{\text{dela}}$ is the delamination width and $w_{\text{coil}} = 14.5 \  \mathrm{mm}$ is the width of the single side of the coil. 
Temperature fields are compared for different width ratios with a fixed delamination thickness of 50 $\mathrm{\mu m}$ in Fig.~\ref{fig:temperature distribution with diff wr}. The results show that a larger width ratio leads to a greater temperature difference across the composite. This also indicates the importance of detecting delamination in the early stages. 

Note that, the 1D forward model used in the inverse model always assumes full-width (global) delamination. Therefore, the influence of the local delamination is reflected in the predicted delamination thickness. This raises an interesting question about the intrinsic correlation between the predicted thicknesses and the actual geometry of the local delamination.

\begin{figure}
    \centering
    \includegraphics[width=0.48\textwidth]{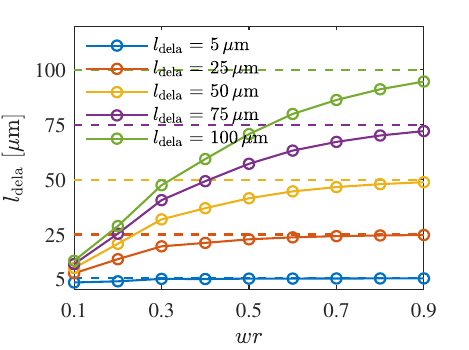}
    \caption{Inverse results of the ten-layer IHCM under local delamination scenario, with $l_{\text{dela}}$ as the unknown. Reference temperature datasets are generated from 2D FE simulations using 45 combinations of preset values: $wr$ ($0.1\sim0.9$) and $l_{\text{dela}}$ (5, 25, 50, 75, 100 $\mathrm{\mu m}$), representing different local delamination conditions.} 
    \label{fig:inverse results dela local_T1}
\end{figure}

By combining the nine width ratios with five thickness values, 45 reference temperature datasets are generated by the 2D FE simulations. These are used to perform inverse modelings, and the results are presented in Fig.~\ref{fig:inverse results dela local_T1}. Where the dashed lines in the figure represent the preset values from the FE simulations. 
As shown, the predicted values reach to the preset values as the width ratio increases. For lower width ratios, the predictions reflect a combined effect of both the width and thickness of the delamination. Results from previous sections has shown that when the width ratio is one (global delamination), the predicted thickness matches the reference line. 

Since both the thickness and width contribute to the measured temperature response, the proposed 1D IHCM cannot resolve them uniquely. Additional constraints are necessary to improve the uniqueness of the solution. Nevertheless, the results exhibit a clear and monotonic relationship with the delamination size, suggesting the potential of the method as an indicator of delamination severity.

\subsection{Case 3: Additional unknowns} \label{additional unknowns}

\begin{figure}
    \centering
    \includegraphics[width=0.48\textwidth]{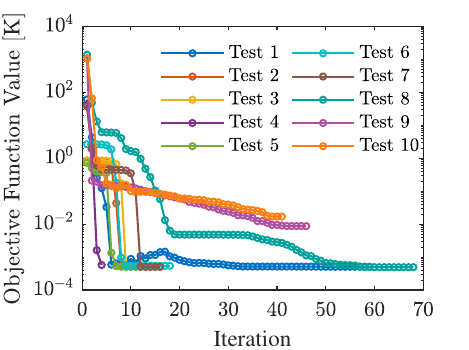}
    \caption{Convergence behaviors of ten test cases, with various combinations of unknowns.} 
    \label{fig:convergence of many unknowns with local_T1}
\end{figure}

In previous sections, the proposed IHCM has been validated by estimating unknown thermal conductivity ($k_{\text{epo}}$), specific heat capacity ($cp_{\text{epo}}$), delamination thickness ($l_{\text{dela}}$), and delamination location. To further evaluate its flexibility and generality, this section explores the IHCM’s capability to handle a broader set of unknown parameters.
Given their significance on thermal and electrical performance, the epoxy layer (i = 3) and Kapton layer (i = 4) are selected for analysis. Their material properties including thermal conductivity ($k_3$, $k_4$), specific heat capacity ($cp_3$, $cp_4$), and density ($\rho_3$, $\rho_4$) are treated as candidate unknowns. In addition, the geometric dimensions of these two layers ($l_3$, $l_4$) and the heat transfer coefficient of the cooling boundary ($h$) are considered.
The analytical forward model is employed within the IHCM, while corresponding 1D FE models are used to generate reference temperature datasets. Different combinations of these parameters are treated as unknowns, requiring tailored IHCMs for each configuration.
The inverse results, along with prediction errors, simulation time, and number of iterations are summarized in Table~\ref{tab:multi_unknown inverse_T1}, and the convergence behaviors for all the test cases are shown in Fig.~\ref{fig:convergence of many unknowns with local_T1}.
\begin{table*}[h!]
\begin{threeparttable}
    \centering
    \caption{Inverse results for various combinations of unknowns, along with their relative errors to preset values.}
    \begin{tabular}{lccccccccccc}
    \hline
    \hline
    & $k_3$ & $k_4$  & $cp_3$ & $cp_4$ & $\rho_3$  &  $\rho_4$   &  $l_3$  &  $l_4$ &  $h$ & Simulation & Number of\\
    \cmidrule(lr){2-3} \cmidrule(lr){4-5} \cmidrule(lr){6-7} \cmidrule(lr){8-9} \cmidrule(lr){10-10}
    Unit & \multicolumn{2}{c}{$\mathrm{[\frac{W}{m \cdot K}]}$} & \multicolumn{2}{c}{$\mathrm{[\frac{J}{kg \cdot K}]}$} & \multicolumn{2}{c}{$\mathrm{[\frac{kg}{m^3}]}$} & \multicolumn{2}{c}{[$\mathrm{mm}$]} & $\mathrm{[\frac{W}{m^2 \cdot K}]}$ & time [$\mathrm{s}$] & iterations \\
    \hline
   \textbf{Preset} &  \textbf{1.3} & \textbf{0.4}  & \textbf{800} & \textbf{1100}  & \textbf{2200} & \textbf{1300} & \textbf{0.2} & \textbf{0.025} & \textbf{1050} \\
    \hline
    Test 1 & 1.294 & 0.404 & & & & & & & &5.3 & 59 \\
    Error [$\%$] & 0.43& 1.03 & & & & & & & \\
    \hline
    Test 2 &  &  & 805 &  995 & & & & & & 1.8 & 16 \\
    Error [$\%$] &  &  & 0.67& 9.53 & & & & & \\
    \hline
    Test 3 &  & & & &2211 & 1202 & & & & 1 & 12 \\
    Error [$\%$] &  & & & &0.5 & 7.51 & & & \\
    \hline
    Test 4 &&&&&&& 0.2 & 0.025&  & 0.5 & 3 \\
    Error [$\%$] &&&&&&& 0.09 & 0.15 & \\
    \hline
    Test 5 & 1.299  & &  & & & & 0.2 & &  & 2 & 12 \\
    Error [$\%$] & 0.09 &  & & & & &0.07 & & \\
    \hline
    Test 6 &   & & 865 & & 2033 & && &  & 1.8 & 18 \\
    Error [$\%$] &  &  & 8.1& & 7.61 & & & & \\
    \hline
    Test 7 & 1.300  & & 800 & &  & && &  & 1.6 & 20 \\
    Error [$\%$] & 0.015 &  & 0.06 & & & & & & \\
    \hline
    Test 8 & 1.31  & & 794 & & & & & & 1049 & 7.4 & 67 \\
    Error [$\%$] & 0.79 &  & 0.75 & & & & & &0.13 \\
    \hline
    Test 9 & 1.665  & & 715 & & 1978 & & 0.211 & & 1021 & 8.8 & 46 \\
    Error [$\%$] & 28.09 &  & 10.63 & & 10.08 & &5.53 & &2.75 \\
    \hline
    Test 10 & 1.247  & 0.541 & 717 & 990 &  & & 0.18 & 0.0125& 1000 & 11 & 40 \\
    Error [$\%$] & 4.09 & 35.15 & 10.33 &10 & & &9.38 &49.82& 4.76 \\
    \hline
    \hline
    \end{tabular}
    \label{tab:multi_unknown inverse_T1}
    \begin{tablenotes}
    \footnotesize
    \item[*] $k_3$, $cp_3$, $\rho_3$, $l_3$ refer to the epoxy layer; $k_4$, $cp_4$, $\rho_4$, $l_4$ refer to the Kapton layer. Error=(predicted value - preset value)/preset value$\times 100$.
    \end{tablenotes}
    \end{threeparttable}
\end{table*}

Several key observations emerge:

1. Same-type thermal property combinations (e.g., $k_3$ and $k_4$, $cp_3$ and $cp_4$, or $\rho_3$ and $\rho_4$ in Tests 1–3): When thermal properties of the same type from different layers are treated as unknowns, compensation effects can occur. One property may be overestimated to offset an underestimated counterpart, meaning that these properties are correlated.

2. Thermally dominant layers yield higher accuracy: The epoxy layer generally provides more accurate inverse results than the Kapton layer. This is due to its greater influence on the measured temperature, while the thinner Kapton layer contributes less and is more susceptible to modeling errors.

3. Combinations of different property types (e.g., ($k_3$, $l_3$), ($k_3$, $cp_3$), or ($k_3$, $cp_3$, $h$) in Tests 5, 7, 8): These configurations produce more accurate results, indicating that combining different types of unknowns improves the robustness of the inversion.

4. Thermal capacitance ambiguity (Test 6): When both $cp$ and $\rho$ are unknown, the IHCM fails to resolve them. Since thermal capacitance is the product of these two parameters, multiple combinations can yield identical temperature profiles. This ambiguity requires additional constraints, such as fixing $\rho$ via separate measurements.

5. All properties of a single material as unknowns (Test 9): This scenario leads to high errors due to strong ill-posedness, indicating the importance of limiting the number of unknowns or incorporating prior knowledge.

6. Most parameters unknown (Test 10): While the overall error increases, predictions for parameters of the thermally dominant layer remain relatively accurate. Less influential parameters are more affected by modeling inaccuracies.

In summary, inverse predictions are more reliable when the unknowns are from layers that are dominate the thermal behavior. Parameters from less influential layers are more sensitive to systematic modeling errors. This aligns with practical interests, where accurate estimation of dominant-layer properties is often more critical. 
This case study demonstrates that the IHCM is applicable to a wide range of unknown combinations, offering flexibility and insight for various multi-layer systems with internal heat generation.

\subsection{Case 4: sensorless approach}

In the previous case studies, the reference temperature was assumed to be measured at the E-E interface using an embedded thermocouple. This is considered an invasive measuring method. In practice, the interface may not be suitable for sensor placement due to space constraints, reliability concerns, or the risk of introducing local defects. Moreover, in high-voltage applications, the presence of a thermocouple may increase insulation complexity. 

To address these limitations, a non-invasive sensorless approach is proposed. This method is suitable for systems where the average temperature can be estimated by non-invasive measurements, such as the current and voltage measurements for coils in electrical machines. Specifically, the coil's transient resistance can be calculated using the measured voltage $U(t)$ and current $I(t)$ by Ohm's law: $R(t) = U(t)/I(t)$.  With the known temperature coefficient of copper, the average temperature of the coil $T_{\text{ave}}(t)$ can be estimated as:
\begin{equation}
    T_{\text{ave}}(t) = T_0 + \beta^{-1} [R(t)/R_0-1],
\end{equation}
where $T_0$ and $R_0$ are the ambient temperature and the coil resistance at that temperature, respectively, and $\beta=3.93\times10^{-3} \ \mathrm{K^{-1}}$ is the temperature coefficient of copper resistivity.

The estimated average temperature $T_{\text{ave}}(t)$ serves as the input to the IHCM in place of direct interface measurements by thermocouple. This is called in this work the sensorless approach, without using thermocouples. In this section, the feasibility of this sensorless approach is evaluated across all the above-investigated inverse scenarios, including unknown material properties, delamination, and various combinations of unknowns.

\subsubsection{Unknown material properties}
\begin{figure*}[ht]
    \centering
    \includegraphics[width=1\linewidth]{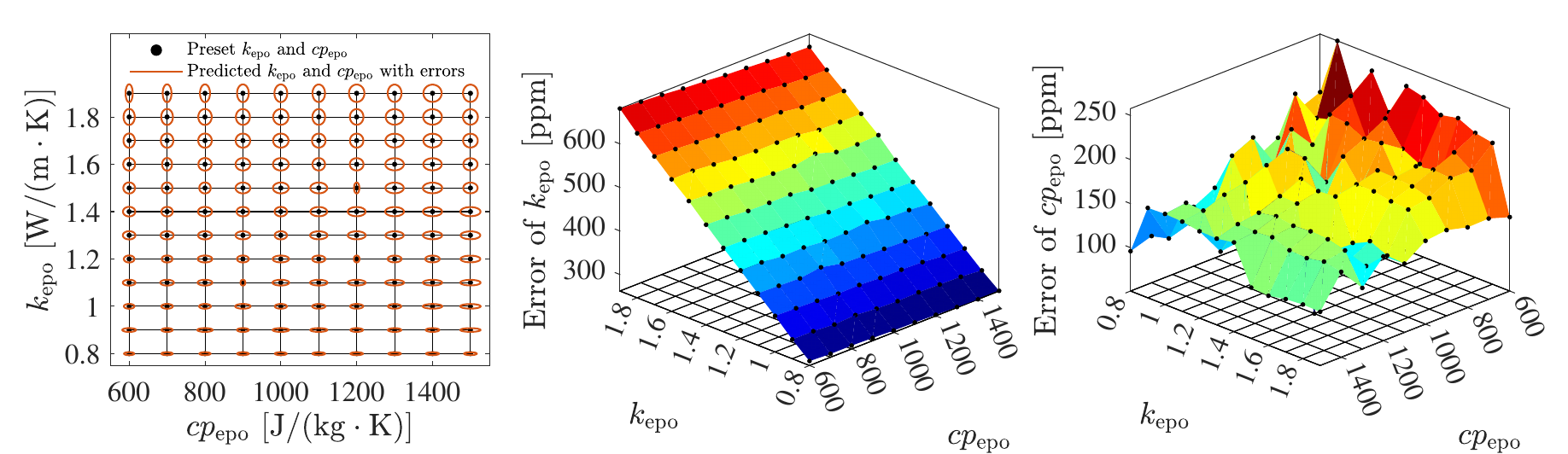}
    \caption{Inverse results of the five-layer sensorless IHCM under a thermal aging scenario, with $k_{\text{epo}}$ and $cp_{\text{epo}}$ as the unknowns. Reference temperature datasets are generated from 1D FE simulations at the E-E interface using 120 combinations of preset values: $k_{\text{epo}}$ (0.8$\sim$1.9 $\mathrm{W/(m \cdot K)}$) and $cp_{\text{epo}}$ (600$\sim$1500 $\mathrm{J/(kg \cdot K)}$), representing varying levels of thermal aging.}
    \label{fig:1D-aging-sensorless}
\end{figure*}
\begin{figure*}[ht]
    \centering
    \includegraphics[width=1\linewidth]{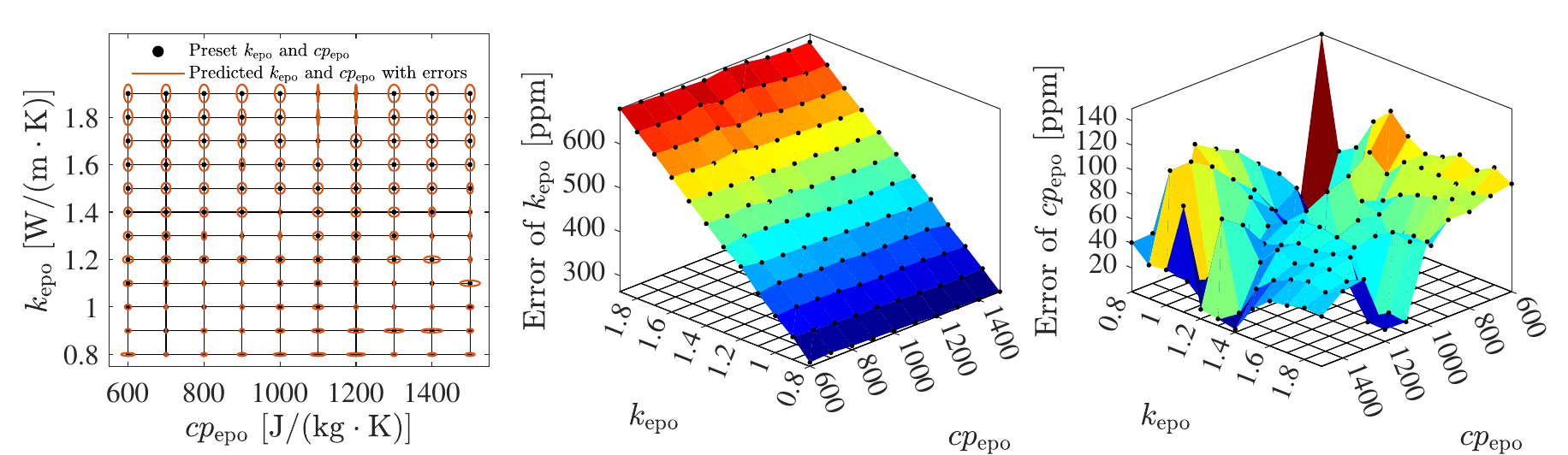}
    \caption{Inverse results of the five-layer sensorless IHCM under a thermal aging scenario, with $k_{\text{epo}}$ and $cp_{\text{epo}}$ as the unknowns. Reference temperature datasets are generated from 2D FE simulations at the E-E interface using 120 combinations of preset values: $k_{\text{epo}}$ (0.8$\sim$1.9 $\mathrm{W/(m \cdot K)}$) and $cp_{\text{epo}}$ (600$\sim$1500 $\mathrm{J/(kg \cdot K)}$), representing varying levels of thermal aging.}
    \label{fig:2D-aging-sensorless}
\end{figure*}

The same 120 combinations of $k_{\text{epo}}$ (0.8$\sim$1.9 $\mathrm{W/(m \cdot K)}$) and $cp_{\text{epo}}$ (600$\sim$1500 $\mathrm{J/(kg \cdot K)}$) used in Section~\ref{section: thermal aging} are considered. Instead of using interface temperature measurements, the IHCM is applied with $T_{\text{ave}}(t)$ as the reference temperature dataset. The corresponding inverse results are presented in Fig.~\ref{fig:1D-aging-sensorless} and Fig.~\ref{fig:2D-aging-sensorless}. 
The results show similar trends in the prediction of $k_{\text{epo}}$, while the error distribution for $cp_{\text{epo}}$ differs slightly due to systematic differences between the 1D and 2D FE models. Compared to the invasive method, the sensorless approach yields slightly higher errors for $k_{\text{epo}}$ but improved accuracy for $cp_{\text{epo}}$, demonstrating the feasibility of this approach in assessing thermal aging.

\subsubsection{Unknown geometric dimensions}
\begin{figure}
    \centering
    \includegraphics[width=0.48\textwidth]{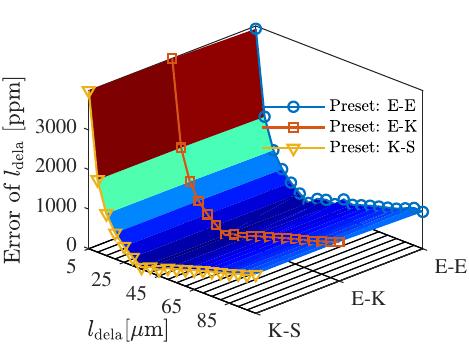}
    \caption{Inverse results of the ten-layer sensorless IHCM under a global delamination scenario, with delamination location and $l_{\text{dela}}$ as the unknowns. Reference temperature datasets are generated from 2D FE simulations using 60 combinations of preset values: three delamination locations (E-E, E-K, and K-S interfaces) and $l_{\text{dela}}$ ($5\sim100\ \mathrm{\mu m}$), representing different global delamination conditions.} 
    \label{fig:inverse results dela global_Tave}
\end{figure}

Delamination scenarios are again considered, using asymmetrical ten-layer models and $T_{\text{ave}}(t)$ as input.
The inverse results for global delamination are shown in Fig.~\ref{fig:inverse results dela global_Tave}.
Consistent with the trend observed in Fig.~\ref{fig:inverse results dela global}, the largest error occurs when the $l_{\text{dela}}$ is 5 $\mu m$, while the error increases as the $l_{\text{dela}}$ exceed 30 $\mathrm{\mu m}$. Notably, the maximum error is approximately four times larger than that in Fig.~\ref{fig:inverse results dela global}. Nevertheless, the delamination locations are accurately identified in all 60 inverse cases, and the highest error is around 4000 ppm (0.4 $\%$), which remains within an acceptable accuracy range.

\begin{figure}
    \centering
    \includegraphics[width=0.48\textwidth]{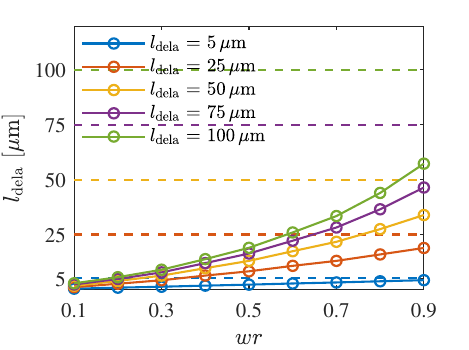}
    \caption{Inverse results of the ten-layer sensorless IHCM under local delamination scenario, with $l_{\text{dela}}$ as the unknown. Reference temperature datasets are generated from 2D FE simulations using 45 combinations of preset values: $wr$ ($0.1\sim0.9$) and $l_{\text{dela}}$ (5, 25, 50, 75, 100 $\mathrm{\mu m}$), representing different local delamination conditions.} 
    \label{fig:inverse results dela local_Tave}
\end{figure}

For local delamination, 2D FE models generate reference datasets $T_{\text{ave}}(t)$ using different combinations of $wr$ and $l_{\text{dela}}$, which are inverted through the ten-layer IHCM. Results are shown in Fig.~\ref{fig:inverse results dela local_Tave}.
The predicted $l_{\text{dela}}$ increases with $wr$, consistent with earlier findings in Fig.~\ref{fig:inverse results dela local_T1}. However, the inverted $l_{\text{dela}}$ values are generally lower than those in Fig.~\ref{fig:inverse results dela local_T1}. This difference arises because the sensorless approach relies on the average coil temperature, which reflects the global heat flux affected by the delamination layer. As a result, it provides a broader but less localized view of the delamination's impact. To quantify the extent of local delamination, a correlation between the inverted and preset $l_{\text{dela}}$ may be established.

\subsubsection{Additional unknowns}
\begin{figure}
    \centering
    \includegraphics[width=0.48\textwidth]{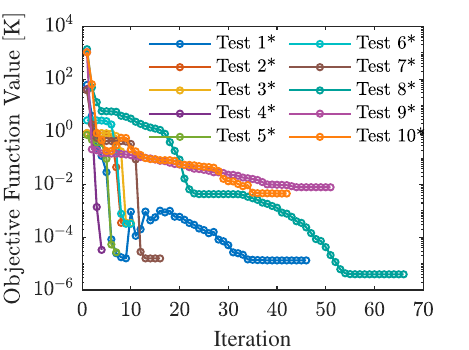}
    \caption{Convergence behaviors of ten test cases (sensorless approach), with various combinations of unknowns.} 
    \label{fig:convergence of many unknowns with local_Tave}
\end{figure}

The sensorless approach is also applied to the same cases as in Section~\ref{additional unknowns} with diverse unknowns, where different combinations of various unknowns are predicted using the five-layer IHCM. The results are summarized in Table~\ref{tab:multi_unknown inverse_Tave}. Compared to Table~\ref{tab:multi_unknown inverse_T1}, the sensorless approach yields similar performance in terms of simulation time and convergence. In some test cases, it even provides higher accuracy, especially when handling more unknowns, such as in Test 10*, where errors in $k_4$ and $l_4$ are notably reduced. This improvement is likely due to the average effect, which reduces sensitivity to local modeling discrepancies. Convergence curves in Fig.~\ref{fig:convergence of many unknowns with local_Tave} confirm that all tests converge successfully. In general, increasing the number of unknowns tends to require more iterations during the inversion process. However, there is no clear monotonic relationship between the number of unknowns and the number of iterations needed for convergence.

\begin{table*}[h!]
    \begin{threeparttable}
    \centering
    \caption{Inverse results of sensorless approach for various combinations of unknowns, along with their relative errors to preset values.}
    \begin{tabular}{lccccccccccc}
    \hline
    \hline
    & $k_3$ & $k_4$  & $cp_3$ & $cp_4$ & $\rho_3$  &  $\rho_4$   &  $l_3$  &  $l_4$ &  $h$ & Simulation & Number of \\
    \cmidrule(lr){2-3} \cmidrule(lr){4-5} \cmidrule(lr){6-7} \cmidrule(lr){8-9} \cmidrule(lr){10-10}
    Unit & \multicolumn{2}{c}{$\mathrm{[\frac{W}{m \cdot K}]}$} & \multicolumn{2}{c}{$\mathrm{[\frac{J}{kg \cdot K}]}$} & \multicolumn{2}{c}{$\mathrm{[\frac{kg}{m^3}]}$} & \multicolumn{2}{c}{[$\mathrm{mm}$]} & $\mathrm{[\frac{W}{m^2 \cdot K}]}$ & time [$\mathrm{s}$] & iterations\\
    \hline
   \textbf{Preset} &  \textbf{1.3} & \textbf{0.4}  & \textbf{800} & \textbf{1100}  & \textbf{2200} & \textbf{1300} & \textbf{0.2} & \textbf{0.025} & \textbf{1050} \\
    \hline
    Test 1* & 1.303 & 0.398 & & & & & & & &3.7 & 46 \\
    Error [$\%$] & 0.23& 0.46 & & & & & & & \\
    \hline
    Test 2* &  &  & 807 &  995 & & & & & & 0.6 & 8 \\
    Error [$\%$] &  &  & 0.97& 9.52 & & & & & \\
    \hline
    Test 3* &  & & & &2218 & 1203& & & & 0.7 & 9 \\
    Error [$\%$] &  & & & &0.82 & 7.45 & & & \\
    \hline
    Test 4* &&&&&&& 0.2 & 0.0249&  & 0.5 & 3 \\
    Error [$\%$] &&&&&&& 0.14 & 0.45 & \\
    \hline
    Test 5* & 1.3  & &  & & & & 0.2 & &  & 1.6 & 7 \\
    Error [$\%$] & 0.02 &  & & & & &0.03 & & \\
    \hline
    Test 6* &   & & 867 & & 2034 & && &  & 0.7 & 9 \\
    Error [$\%$] &  &  & 8.39 & & 7.57 & & & & \\
    \hline
    Test 7* & 1.301  & & 800 & &  & && &  & 1.4 & 16 \\
    Error [$\%$] & 0.05 &  & 0.02 & & & & & & \\
    \hline
    Test 8* & 1.31  & & 795 & & & & & & 1049 & 7.2 & 66 \\
    Error [$\%$] & 0.74 &  & 0.58 & & & & & &0.11 \\
    \hline
    Test 9* & 1.665  & & 715 & & 1979 & & 0.211 & & 1022 & 9.9 & 51 \\
    Error [$\%$] & 28.09 &  & 10.51 & & 10.07 & &5.74 & &2.72 \\
    \hline
    Test 10* & 1.399  & 0.486 & 721 & 990 &  & & 0.195 & 0.0253& 1024 & 11.6 & 42 \\
    Error [$\%$] & 7.6 & 21.54 & 9.88 &9.99 & & &2.29 &1.22&2.5 \\
    \hline
    \hline
    \end{tabular}
    \label{tab:multi_unknown inverse_Tave}
    \begin{tablenotes}
    \footnotesize
    \item[*] $k_3$, $cp_3$, $\rho_3$, $l_3$ refer to the epoxy layer; $k_4$, $cp_4$, $\rho_4$, $l_4$ refer to the Kapton layer. Error=(predicted value - preset value)/preset value$\times 100$.
    \end{tablenotes}
    \end{threeparttable}
\end{table*}

\subsubsection{Summary}
A non-invasive sensorless temperature measuring approach for the IHCM is proposed and validated across several inverse scenarios, including thermal aging, delamination, and diverse combinations of unknown parameters. The results indicate that this method achieves comparable accuracy to traditional thermocouple-based approaches while avoiding physical intrusiveness and mitigating measurement errors.
The sensorless approach is especially promising for high-voltage or compact systems, where deploying conventional sensors are impractical.
However, since the average temperature smooths out local variations, it may miss localized defects.
Experimental validation is necessary to assess performance under real-world conditions, where additional factors such as natural air convection, radiation, and unstable cooling system can influence results.
Notably, the current 1D IHCMs can be extended into 2D or 3D to capture more complex geometries and multidimensional heat flow. It will enhance spatial resolution and enable detection of localized anomalies, while increases computational cost.
\section{Conclusions}

This paper introduces a fast and accurate inverse heat conduction modeling approach for the parameter estimation in multi-layer composites with internal heat generation, and concludes the second part of this two-part study. The inverse models are constructed using analytical 1D forward solutions based on the SOV-OE and Green's function methods, which were developed and validated in the first paper.

The proposed IHCM is applied to various practical scenarios. A five-layer configuration is used to study the effects of thermal aging, material degradation, and manufacturing tolerances, demonstrating the model’s ability to accurately estimate variations in material properties. A ten-layer configuration is employed to investigate global and local delamination, where both the location and thickness of the delaminated layer are successfully identified. Additionally, ten test cases with different combinations of unknowns are examined using the five-layer IHCM, demonstrating both its versatility and its limitations.
A sensorless temperature measurement approach is also proposed. By estimating the coil’s average temperature through electrical measurements, this method supports non-invasive applications and reduces measurement-related systematic errors.

Overall, this inverse modeling framework provides a practical and computationally efficient tool for identifying internal defects and property changes in multi-layer composites subjected to internal heat generation. It holds strong potential for integration into lightweight, portable systems for online thermal monitoring and fault prognosis.

\printcredits

\bibliographystyle{elsarticle-num_nourl}

\bibliography{References}

\end{document}